\documentclass[aps,prb,twocolumn,floatfix,groupedaddress,showpacs]{revtex4}
\usepackage{graphicx}
\usepackage{dcolumn}
\usepackage{bm}
\begin{document}

\title{Effect of electron and hole doping on
       the structure of C, Si, and S nanowires}

\author{Shinya Okano}
\affiliation{Physics and Astronomy Department,
             Michigan State University,
             East Lansing, Michigan 48824-2320}

\author{David Tom\'anek}
\email[E-mail: ]{tomanek@msu.edu}
\affiliation{Physics and Astronomy Department,
             Michigan State University,
             East Lansing, Michigan 48824-2320}

\date{\today}

\begin{abstract}
We use {\em ab initio} density functional calculations to study
the effect of electron and hole doping on the equilibrium geometry
and electronic structure of C, Si, and S monatomic wires.
Independent of doping, all these nanowires are found to be
metallic. In absence of doping, C wires are straight, whereas Si
and S wires display a zigzag structure. Besides two preferred bond
angles of $60^\circ$ and $120^\circ$ in Si wires, we find an
additional metastable bond angle of $90^\circ$ in S wires. The
equilibrium geometry and electronic structure of these nanowires
is shown to change drastically upon electron and hole doping.
\end{abstract}



\maketitle



The close relationship between geometric and electronic structure
of molecules and solids underlies the richness of phenomena
described by Physics and Chemistry. The electronic structure plays
a central role in determining the equilibrium geometry by
minimizing the corresponding free energy. Geometrical structure,
in turn, determines the electrostatic field confining the
electrons. This subtle two-way relationship between geometry and
electronic structure underlies the complex behavior of even simple
systems, such as monatomic nanowires or triatomic molecules. In
the latter case, rather heuristic rules, such as the Walsh
rule\cite{walsh53}, have been used to rationalize, whether a
trimer carrying a particular number of electrons should be linear
or bent. Approaches based on the theory of directed valence,
including the Valence Shell Electron Pair Repulsion (VSEPR)
model\cite{gillespie}, have generally proven useful in determining
the bonding geometry in polyatomic molecules. These approaches,
however, miss subtle differences in bonding between elements of
the same group and are too rough to predict the effect of doping
on the bonding geometry.

Most previous studies have focussed on elements with a ductile,
metallic bulk phase due to their capability to form nanowires by
mechanical stretching. Interest in the geometry of such nanowires
has been triggered by electron microscopy observations of
monatomic Au nanowires with atomic resolution, and the correlation
between their atomic structure and quantum
conductance.\cite{ohnishi98:_quant} Even the structure of the
simplest nanowires is still a matter of debate, since the observed
inter-atomic distance was found to be significantly larger than
the bulk inter-atomic distances. Several explanations have been
offered as an interpretation of the initial experimental data.
These include a rotating zigzag wire
structure,\cite{sanchez-portal00:_zigzag} stabilization of the
nanowire by impurity atoms with a low cross-section for
electrons,\cite{legoas02:_origin,novaes03:_effec} and charging
effects that would expand the inter-atomic
distance.\cite{ayuela05:_charg}

Inspired by these investigations on Au, theoretical studies have
been expanded to nanowires of other metals, including Al, Ag, Pd,
Rh, and Ru.\cite{ribeiro03:_au} Some of these nanowires are
reported to exhibit zigzag deformations with more than one stable
bond angle at zero strain. In contrast to VSEPR model predictions,
theoretical studies of group IV elements suggest that C forms a
linear wire, whereas the Si forms a zigzag
wire.\cite{senger05:_atomic,tongay05:_atomic} Nanowires of
non-metals are of interest as either free-standing systems, or
constituents of molecular electronics
devices.\cite{joachim00:_elect} In the latter case, their
geometrical and electronic structure, and thus their transport
properties, may be influenced by charge doping caused by the
chemical environment and the presence of a current.

In the present study, we use {\em ab initio} density functional
calculations to investigate the effect of charge doping and
structural constraints on the equilibrium geometry and electronic
structure of C, Si, and S monatomic wires. For different levels of
doping, we present total energy plots as a function of lattice
constant and bond angle, which provide important information about
the structure and stiffness of these nanowires, when exposed to
particular environments or deformations. We find neutral C wires
to be straight, whereas Si and S wires display a zigzag structure.
Besides two preferred bond angles of $60^\circ$ and $120^\circ$ in
Si wires, we find an additional metastable bond angle of
$90^\circ$ in S wires. We also observe changes in the equilibrium
geometry and electronic structure induced by doping, including a
zigzag distortion of electron doped C nanowires and the number of
stable geometries in S nanowires depending on the doping level.

Our calculations for C, Si, and S monatomic chains are based on
the density functional theory (DFT)\cite{HK,KS} within the local
density approximation (LDA). We use Troullier-Martins {\em ab
initio} pseudopotentials to describe the interaction of valence
electrons with atomic cores\cite{TM} and the Perdew-Zunger form of
the exchange-correlation potential\cite{CA,PZ81}, as implemented
in the SIESTA code.\cite{SIESTA1,SIESTA2} Our basis consists of
double-zeta localized orbitals with polarization functions (DZP).
The range of the localized orbitals is limited in such a way that
the energy shift caused by their spatial confinement is no more
than 100~meV.\cite{{sankey89},{SIESTA_PAO},{Eshift-convergence}}
The charge density and potentials are calculated on a real-space
grid with a mesh cutoff energy of $200$~Ry for C and $150$~Ry for
Si and S. This is sufficient to achieve a total energy convergence
of $1$~meV/atom. We use periodic boundary conditions in all
directions and separate the nanowires laterally by $9$~{\AA} to
prevent them from interacting. We sample the 1D Brillouin zone of
the nanowires by $96$~k-points for all the lattice constants
considered. Since electronic structure calculations in a
superlattice geometry require global charge neutrality, we perform
calculations for charged chains on the background of a uniformly
distributed counter-charge.

\begin{figure}[t]
\includegraphics[width=0.8\columnwidth]{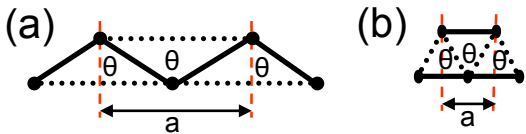}
\caption{(Color online) Schematic of a unit cell of (a) a
single-stranded and (b) double-stranded nanowire, with the unit
cell of length $a$ delimited by the dashed lines. The key
distinguishing feature between the two structures is the value of
the bond angle $\theta$. The atomic positions are indicated by
solid circles, nearest neighbor bonds by solid lines, second
neighbor bonds by dashed lines. \label{fig1}}
\end{figure}

For a given lattice constant, we globally optimize each system to
determine its optimum geometry, total energy, and electronic
structure. We accommodate two atoms per unit cell, and use an
initial geometry with two equivalent bonds as a starting point of
the optimization. We identify each nanowire structure by the pair
of values $(a,{\theta})$, where $a$ is the lattice constant in
{\AA}ngstroms and $\theta$ the bond angle in degrees, as depicted
in Fig.~\ref{fig1}(a). Obviously, for bond angle values
$\theta{\alt}60^\circ$, the second neighbor interaction along the
wire direction may dominate over the nearest neighbor interaction,
causing a spontaneous transition to a double-chain. This geometry
is depicted in Fig.~\ref{fig1}(b). We consider a structure to be
optimized when none of the residual forces acting on atoms exceed
the value of $0.01$~eV/{\AA}.

\begin{figure}[bt]
\includegraphics[width=0.7\columnwidth]{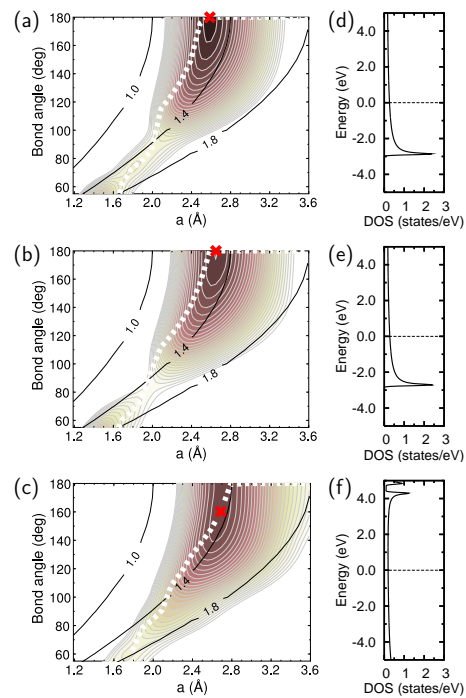}
\caption{(Color online) Energy of a C nanowire as a function of
the lattice constant $a$ and the bond angle, for an undoped system
(a), and for doping levels of $+0.5$~electrons/atom (b), and
$-0.5$~electrons/atom (c). Contour lines, representing the total
energy per unit cell, are separated by $0.2$~eV. Optimized
geometries for each lattice constant are shown by the white dotted
lines. Lines of constant bond length are indicated by black solid
lines. The electronic density of states for the globally optimized
geometries, denoted by the solid $\times$ in (a-c), are shown in
(d-f), respectively. The Fermi level lies at $E=0$. \label{fig2}}
\end{figure}

Fig.~\ref{fig2} shows contour maps of the total energy of carbon
nanowires $E_{tot}(a,{\theta})$ as a function of the lattice
constant $a$ and bond angle $\theta$. Independent of doping, we
generally find the bond angle to increase with increasing lattice
constant. Especially in systems where the bond length is stretched
beyond the equilibrium value, this trend reflects the tendency to
optimize the bond length first and only then the bond angle,
suggesting that bond stretching is harder than bond bending.

For a neutral carbon wire, depicted in Fig.~\ref{fig2}(a), the
equilibrium structure is found at $(2.6,180)$, corresponding to a
linear chain. When the bond angle is constrained to
$\theta{\alt}60^\circ$, a secondary minimum evolves at $(1.4,20)$,
corresponding to a double-chain with only a slightly longer
nearest neighbor distance in each strand. For a hole-doped wire
carrying $+0.5$~electrons/atom, depicted in Fig.~\ref{fig2}(b),
the optimum structure is still linear and the interatomic spacing
is unchanged, corresponding to $(2.6,180)$. In contract to the
undoped system, however, the Coulomb repulsion suppresses the
secondary minimum corresponding to a paired wire configuration.
Upon electron doping at the level of $-0.5$~electrons/atom,
depicted in Fig.~\ref{fig2}(c), the global geometry optimum shows
a much weaker dependence on the bond angle than in neutral or hole
doped wires. The system gains ${\approx}30$~meV/atom, when
relaxing from the linear structure at $(2.7,180)$ to a bent
geometry, characterized by $(2.6,150)$. Increased electron doping
up to $-0.7$~electrons/atom further changes the bond angle to
$\theta{\approx}140^\circ$ and the bond length to $1.47$~{\AA} at
a still larger energy gain with respect to the linear structure
with $1.3$~{\AA} as optimum bond length. Also in electron doped
carbon nanowires, the secondary minimum corresponding to a paired
wire configuration is suppressed. Even though the interaction
between neutral wires is negligible at the inter-wire separation
of $9$~{\AA}, the Coulomb repulsion between charged wires is
strong enough to suppress the secondary minima associate with
paired wires. This has been confirmed by observing a stabilization
of charged nanowire systems when increasing the inter-wire
separation.

The robustness of the linear structure of carbon chains is also
reflected in the electronic density of states for the globally
optimized geometries. As seen in Figs.~\ref{fig2}(d-f), the
density of states is rather featureless, suggesting that electron
or hole doping does not modify the nature of electronic states
responsible for bonding. Whether neutral or doped, carbon
nanowires are metallic, with a nearly constant density of states
near the Fermi level. The van Hove singularities in the spectrum
are characteristic of one-dimensional systems.

\begin{figure}[t]
\includegraphics[width=0.7\columnwidth]{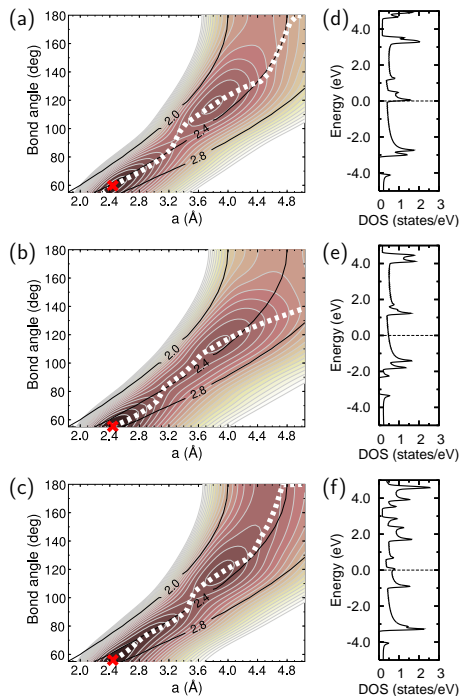}
\caption{(Color online) Energy of a Si nanowire as a function of
the lattice constant $a$ and the bond angle, for an undoped system
(a), and for doping levels of $+0.5$~electrons/atom (b), and
$-0.5$~electrons/atom (c). Contour lines, representing the total
energy per unit cell, are separated by $0.2$~eV. Optimized
geometries for each lattice constant are shown by the white dotted
lines. Lines of constant bond length are indicated by black solid
lines. The electronic density of states for the globally optimized
geometries, denoted by the solid $\times$ in (a-c), are shown in
(d-f), respectively. The Fermi level lies at $E=0$. \label{fig3}}
\end{figure}

\begin{figure}[t]
\includegraphics[width=0.7\columnwidth]{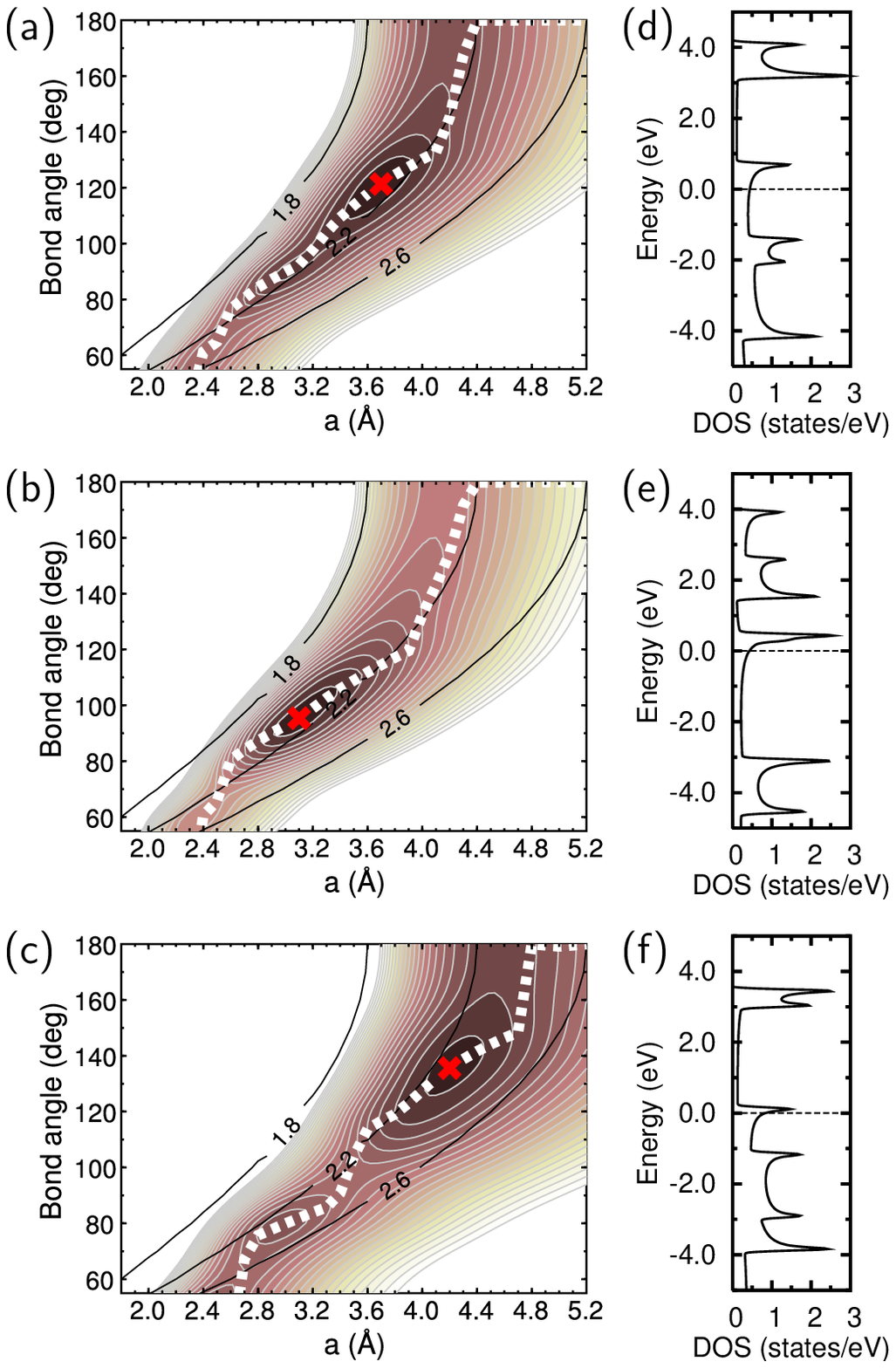}
\caption{(Color online) Energy of an S nanowire as a function of
the lattice constant $a$ and the bond angle, for an undoped system
(a), and for doping levels of $+0.5$~electrons/atom (b), and
$-0.5$~electrons/atom (c). Contour lines, representing the total
energy per unit cell, are separated by $0.2$~eV. Optimized
geometries for each lattice constant are shown by the white dotted
lines. Lines of constant bond length are indicated by black solid
lines. The electronic density of states for the globally optimized
geometries, denoted by the solid $\times$ in (a-c), are shown in
(d-f), respectively. The Fermi level lies at $E=0$. \label{fig4}}
\end{figure}

\begin{figure}[t]
\includegraphics[width=\columnwidth]{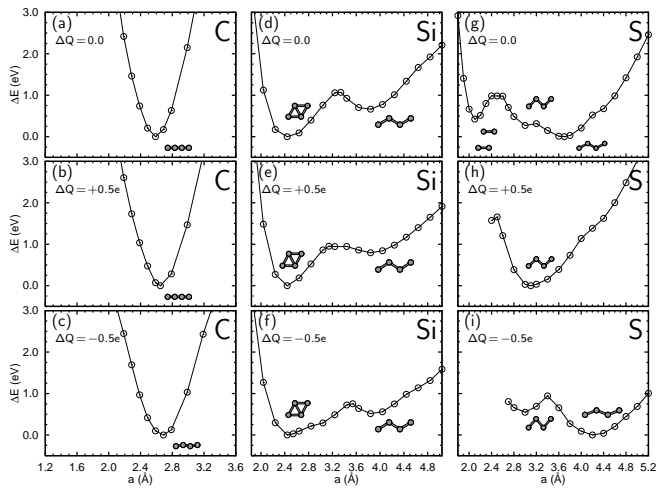}
\caption{Total energy changes in neutral and doped monatomic
nanowires of C, Si and S as a function of the imposed lattice
constant $a$, with the bond angle $\theta(a)$ given by the white
dotted lines in Fig.~\protect\ref{fig2}(a-c) for C,
Fig.~\protect\ref{fig3}(a-c) for Si, and
Fig.~\protect\ref{fig4}(a-c) for S. \label{fig5}}
\end{figure}

Contour maps of the total energy $E_{tot}(a,{\theta})$ of silicon
nanowires are shown in Fig.~\ref{fig3}. The potential energy
surface of undoped Si nanowires exhibits two minima at $(3.8,120)$
and $(2.4,60)$, as shown in Fig.~\ref{fig3}(a). The bond length
$d{\approx}2.2$~{\AA} in the structure with
$\theta{\approx}120^\circ$ is shorter than in the double-stranded
structure, formed by equilateral triangles with
$d{\approx}2.5$~{\AA}. As seen in Figs.~\ref{fig3}(b-c), the shape
of the energy surface changes only slightly upon doping. In
particular, the two stable geometries remaining nearly unchanged,
but the one with the smaller bond angle becomes more stable upon
doping.

Even though both silicon and carbon are group IV elements, the
geometries of these nanowires are very different. The underlying
cause is the different electronic structure of these nanowires,
reflected in the electronic density of states for the globally
optimized geometries. Same as C chains, also Si chains maintain a
metallic character. In contrast to Figs.~\ref{fig2}(d-f), the
density of states of Si nanowires in Figs.~\ref{fig3}(d-f) is very
rich in features, suggesting that the equilibrium geometry is
largely determined by the energetic availability of particular
orbitals for bonding.

As suggested previously~\cite{PhysRevLett.57.1145} and seen in
Fig.~\ref{fig4}(a), the potential energy surface of S nanowires is
more structured than that of C and Si nanowires. The global
minimum in the neutral system is at $a=3.6$~{\AA} and
$\theta=120^\circ$. Similar to C and Si nanowires, we find a
second shallow minimum near $\theta=60^\circ$. In addition to the
$\theta=120^\circ$ and $60^\circ$ bond angles, found in other
nanowires, the S wire shows an additional minimum at
$\theta=90^\circ$. Not shown in Fig.~\ref{fig4}(a) is the fourth
minimum at $(2.2,40)$, corresponding to a pair of linear wires
with the bond length $d=2.2$~{\AA}. The bond length in the zigzag
wires with $\theta=90^\circ$ and $120^\circ$ is $d=2.1$~{\AA} and
increases to $2.4$~{\AA} at $\theta=60^\circ$. In contrast to Si
wires, doping by electrons and holes changes the total energy
surface of S wires drastically. The system carrying a net charge
$+0.5$~electrons/atom, depicted in Fig.~\ref{fig4}(b), has only
one minimum at $a=3.0$~{\AA} and $\theta=90^\circ$. There is also
an indication of a second spurious minimum at $a=2.2$~{\AA} and
$\theta<60^\circ$, indicating the preference of the system in a
short unit cell to dissociate into two parallel wires that repel
each other. In the electron-doped system at the level of
$-0.5$~electrons/atom, depicted in Fig.~\ref{fig4}(c), we observe
two minima at $(4.2,135)$ and $(3.0,80)$. There is also a spurious
additional minimum near $(2.2,27)$, an indication of the
instability of a charged S nanowire, crowded in a short unit cell,
with respect to two parallel wires subject to Coulomb repulsion.

The electronic density of states of S chains, depicted in
Figs.~\ref{fig4}(d-f), is very rich in features, same as that of
Si nanowires in Figs.~\ref{fig3}(d-f), thus providing a background
for explaining the number of stable morphologies and their
deviation from heuristic models. Similar to C and Si chains, the
electronic spectrum of sulfur nanowires shows no gaps near the
Fermi level, indicating metallic behavior.

Total energy differences in neutral and doped C, Si and S
nanowires are shown in Fig.~\ref{fig5} as a function of the
lattice constant $a$ along the optimum trajectories $\theta(a)$,
given by the white dotted lines in
Figs.~\ref{fig2}-\ref{fig4}(a-c). Clearly visible is the number of
energy minima, which changes from element to element. A second,
clear message emerges, namely that electron or hole doping plays a
much less significant role in carbon nanowires than in those of
silicon and, even more so, of sulfur.

For the charge neutral systems, the equilibrium bond angles in
nanowires follow the trends found in trimers. Both the C trimer
and nanowire are linear, whereas the Si trimer and nanowire are
bent, with the bond angles of ${\approx}80^\circ$ in the trimer
and the secondary minimum at ${\approx}120^\circ$ in the Si
nanowire. Similar to the S nanowire, the S trimer exhibits a more
complex energy surface with two minima at ${\theta}=120^\circ$ and
$60^\circ$. The structural difference between C$_3$ and Si$_3$ had
been formerly partly attributed to the role of $d$ orbitals in
Si$_3$.\cite{jones85} Our results differ from these findings,
since Si$_3$ bends spontaneously even when the $d$ orbitals are
absent.\cite{Si-basis}

Even though the initial nanowire geometry, characterized by
$(a,\theta)$, had two equivalent bonds per unit cell, the
optimizations have been performed with no further constraints. We
found that also all our final structures have two equivalent
bonds, but cannot exclude the possibility of more stable
structures with inequivalent bond lengths. In the present work, we
confined ourselves to finding out, whether particular wires are
linear or bent. Even though bond length alternation caused by a
Peierls instability is an interesting problem, its exhaustive
study exceeds the scope of the present work. Reliable results are
not easy to come by, since calculated Peierls distortions appear
to partly depend also on the computational approach, with
distortions predicted by LDA/GGA generally smaller than those
based on Hartree-Fock
calculations.\cite{abdurahman00,tongay05:_atomic}

In summary, we used {\em ab initio} density functional
calculations to investigate the effect of charge doping and
structural constraints on the equilibrium geometry and electronic
structure of C, Si, and S monatomic wires. Unlike bulk C, Si, and
S, the nanowires of these elements are found to be metallic. For
different levels of electron and hole doping, we determined the
total energy as a function of lattice constant and bond angle.
These results provide important information about the equilibrium
structure and stiffness of the nanowires, when exposed to
particular environments or deformations. We found neutral C wires
to be straight, whereas Si and S wires displayed a zigzag
structure. Besides two preferred bond angles of $60^\circ$ and
$120^\circ$ in Si wires, we found an additional metastable bond
angle of $90^\circ$ in S wires. We found carbon nanowires to be
linear and rigid, with a tendency to zigzag distortion at moderate
electron doping. Whereas the zigzag geometry of silicon nanowires
is not affected by doping, that of sulfur nanowires changes
strongly, with even the number of stable geometries depending on
the doping level.

This work was partly supported by the NSF NIRT grant ECS-0506309,
the NSF NSEC grant EEC-0425826, and the Humboldt Foundation Award.



\begin{thebibliography}{25}
\bibitem[{Esh()}]{Eshift-convergence}
\bibinfo{note}{We found our results for energy differences associated with
  chain distortions to be converged when using $100$~meV as the energy shift
  parameter. Reducing this value to $50$~meV or increasing it to $150$~meV
  changes the relevant energy differences by only $0.01-0.02$~eV per unit
  cell.}
\bibitem[{Si-()}]{Si-basis}
\bibinfo{note}{The equilibrium bond angle in Si$_3$ is modified by only
  ${\approx}20^\circ$ when changing from the DZP (double-$\zeta$ basis for $s$ and
  $p$, single-$\zeta$ basis for $d$) to the SZ (single-$\zeta$ basis for $s$ and $p$,
  no $d$ orbitals) basis set.}
\end{thebibliography}

\end{document}